\pdfoutput=1 % helps arXiv pick pdfLaTeX

\documentclass [
  acmsmall,
  screen,
  nonacm
] {acmart}

\usepackage [utf8] {inputenc}

\usepackage {
  array,
  booktabs,
  cleveref,
  framed,
  graphicx,
  kantlipsum,
  listings,
  nicefrac,
  pdflscape,
  tcolorbox,
  tikz,
  xltabular,
}

\usetikzlibrary {arrows.meta,decorations.pathreplacing}
\usepackage [export]             {adjustbox}
\usepackage [inline,shortlabels] {enumitem}
\usepackage [table]              {xcolor}

\lstdefinestyle {pymc} {
  language          = Python,
  basicstyle        = \ttfamily\footnotesize,
  commentstyle      = \color{gray!80!black}\itshape,
  keywordstyle      = \bfseries,
  showstringspaces  = false,
  columns           = fullflexible,
  keepspaces        = true,
  escapeinside      = {(*}{*)},
  literate          = {rho}{{$\rho$}}1 {eta}{{$\eta$}}1 {pi}{{$\pi$}}1,
}

\crefname {figure}  {Fig.}    {figures}
\Crefname {figure}  {Figure}  {Figures}
\crefname {section} {Sect.}   {sections}
\Crefname {section} {Section} {Sections}
\crefname {table}   {Tbl.}    {tables}
\Crefname {table}   {Table}   {Tables}

\newcolumntype P [1]
  {>{\raggedright \arraybackslash}p{#1}}

\newcolumntype R [1]
  {>{\raggedleft \arraybackslash}p{#1}}

\newenvironment {rqconclusion}
  {\begin {leftbar}\noindent \ignorespaces}
  {\end {leftbar}}

\theoremstyle{acmdefinition}
\newtheorem*{example*}{Example}

\begin {document}

\title [Software Testing Meets Bayesian Workflow]
  {Software Testing Meets Bayesian Workflow:\\A Literature-Driven Artifact Study of Probabilistic Programs}

\author {Oleg Jarma \and Maja Dall \and Ra{\'u}l Pardo \and Mahsa Varshosaz \and Andrzej Wąsowski}

\begin{abstract}

  Automated testing is an effective software quality assurance method, but it is rarely used for probabilistic models. This is despite the heavy use of probabilistic programming for data analytics in research and engineering, where the consequences of errors are dire. A key obstacle is the specification of testable correctness properties for probabilistic models, known as the oracle problem.  Like machine learning, probabilistic programming is used precisely where no analytical specification exists: had we been able to write one, we would have written an algorithm instead of fitting a model. An alternative is partial, proxy correctness specifications. The statistics literature already offers many, but as quality guidelines that require human judgment: comparing plots, visually analyzing shapes of distributions, assessing whether parameter values are ``far'' or ``close enough''. These guidelines are not systematically automated.

  In this work, we elicit, implement, and study testable correctness properties of probabilistic models. We extract 17 broadly applicable properties for probabilistic models from one of the most widely used textbooks in Bayesian data analysis, \emph{Statistical Rethinking} by McElreath. For each property, we select and implement a test oracle, and study its utility for automatic testing, using a population of 59 probabilistic models from academic and industrial sources. We find realistic use cases (models under test) for all the properties.  We observe that the vast majority of tests are not flaky in practice, despite the theoretical potential for flakiness. To the best of our knowledge, we are the first to investigate the effectiveness of mutation testing for probabilistic models, showing that standard mutation operators do not capture faults tracked by the probabilistic properties.  With this work we hope to encourage data analysts to automate their quality assurance practices, and to attract the software engineering research community to building testing infrastructures for probabilistic models.

\end{abstract}

\maketitle

\section {Introduction}%
\label {sec:introduction}

\begin {figure} [t]

  \begin {minipage} {.47 \textwidth}
  \[
    \small
     \begin {array} {
       @{}
       l
       r
       @{~}
       l
       @{}
     }
       \text{prior:}
       & \rho & \sim \mathrm{Uniform}(0, 2000)
       \\[2pt]

       \text{predictor:}
       & \eta_i & = \log\rho - i\log 2, \qquad
       \\[2pt]

       \text{link:}
       & \pi_i & = \mathrm{cloglog}^{-1}\eta_i
       \\[2pt]

       \text{likelihood:}
       & \mathrm{data}_i & \sim \mathrm{Binomial}(N, \pi_i), \quad i = 1,\dots,K
     \end{array}
  \]
  \end {minipage}%
  \hfill%
  \begin {minipage} {.47 \textwidth}
    \scalebox {0.54} {% !TEX root = ../main.tex

\begin{tikzpicture}[
    x=1.3cm, y=1.3cm,                      % = cetz canvas length
    img/.style={inner sep=0pt, outer sep=0pt},
    arr/.style={-{Stealth[length=2mm]}, line width=.4pt},
  ]

  % ---- beakers ------------------------------------------------------
  \foreach \i/\n/\x in {0/1/0.6, 1/2/2.6, 2/3/5.2, 3/4/7.8}
    \node[img] (b\i) at (\x,0) {\includegraphics[width=0.95cm]{figures/beaker\n.pdf}};

  % ---- petri dishes, 2x2 per stage ----------------------------------
  \foreach \s/\a/\b/\c/\d in {1/clean/dirty/dirty/dirty,
                              2/clean/dirty/clean/dirty,
                              3/dirty/clean/clean/clean} {
    \pgfmathsetmacro{\X}{2.6*\s}
    \node[img] (stage\s) at (\X-0.3,-1.5)
      {\includegraphics[width=8mm]{figures/petri-dish-\a.pdf}};

    \node[img]           at (\X+0.4,-1.5)
      {\includegraphics[width=8mm]{figures/petri-dish-\b.pdf}};

    \node[img]           at (\X+0.3,-2.2)
      {\includegraphics[width=8mm]{figures/petri-dish-\c.pdf}};

    \node[img]           at (\X-0.4,-2.2)
      {\includegraphics[width=8mm]{figures/petri-dish-\d.pdf}};
  }

  % ---- initial concentration ----------------------------------------
  \node at ([shift={(0.10,-0.28)}]b0.north east) {\Large$\rho$};

  % ---- braces --------------------------------------------------------
  \draw[decorate, decoration={brace, amplitude=5pt}]
       ([shift={(-0.2,-0.9)}]stage1.south west) --
       ([shift={(-0.2, 0.1)}]stage1.north west)
       node[midway, left=6pt] {\Large$N \, \times$};

  \draw[decorate, decoration={brace, amplitude=5pt}]
       ([shift={(-0.81,0.7)}]b1.north) --
       ([shift={( 0.70,0.7)}]b3.north)
       node[midway, above=6pt] {\Large$K \, \times$};

  % ---- dilution steps ------------------------------------------------
  \foreach \i/\j in {0/1, 1/2, 2/3} {
    \node[img] at ([shift={(-0.05,0.30)}]b\j.north west)
      {\reflectbox{\includegraphics[width=10mm]{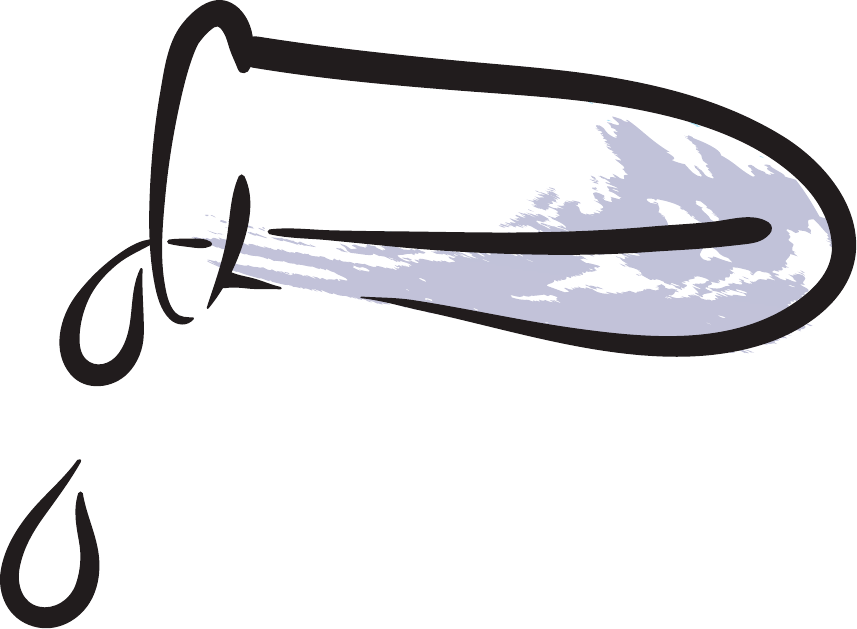}}};

    \node[anchor=west] at ([shift={(-0.10,-0.28)}]b\j.north east)
      {\Large$\rho_\j\,=\,2^{-\j}\!\rho$};

    \draw[arr] ([shift={( 0.14,0)}]b\i.east) -- ([shift={(-0.14,0)}]b\j.west);
    \draw[arr] ([shift={(0,-0.112)}]b\j.south) --
               ([shift={(0.14,0.07)}]stage\j.north east);

    \node at ([shift={(0.10,-0.35)}]b\j.south east) {\Large$\pi_\j$};
    \node at ([shift={(1.12,-0.65)}]stage\j.south east) {\LARGE$\mathrm{data}_\j$};
  }
\end{tikzpicture}
}
  \end {minipage}%

  \vspace {-2ex plus 2pt minus 1.5pt}

\begin {lstlisting} [style = pymc]
with pm.Model():
    rho  = pm.Uniform("rho", 0, 2000)                     (*$\leftarrow$ prior*)
    eta  = pm.math.log(rho) - np.arange(1, K+1)*np.log(2) (*$\leftarrow$ predictor*)
    pi  = 1 - pm.math.exp(-pm.math.exp(eta))             (*$\leftarrow$ link*)
    pm.Binomial("data", n=N, p=pi, observed=data)      (*$\leftarrow$ likelihood*)
\end{lstlisting}

\vspace {-1ex plus 2pt minus 1pt}

\caption {Top: Fisher's dilution assay model\,\cite {fisher:1922}, a formulation as a GLM following McCullagh and Nelder\,\cite {mccullagh.nelder:1989}.  We estimate a concentration $\rho$ of a contaminant in a liquid across $K$ dilution steps of $N$ tests each.  Bottom: An implementation of the model as a Bayesian probabilistic program in PyMC, an internal DSL of Python.}%
\label {fig:example}

\vspace {-1ex plus 3pt minus 1pt}

\end {figure}

Computational methods have spread across research fields from humanities, social science, engineering, biology, ecology, to physics~\cite{mesr:2024:bso}.  Many, if not most, are concerned with statistical data analysis using \emph{probabilistic models} (\cref {fig:example}) implemented in probabilistic programming languages\,\cite {gilks.ea:1994:bugs,carpenter.ea:2017:stan,abril-pla.ea:2023:pymc,bingham.ea:2019:pyro,plummer:2003:jags,ge.ea:2018:turing}, often internal domain-specific languages (DSLs) embedded in a general purpose programming language. The use of software enables important breakthroughs but it is also a source of problems. For instance, in 2014,  Lin et al. published an analysis of~gene expression, measuring how active different genes are\,\cite {lin.ea:2014:comparison}. A year later Gilad and Mizrahi-Man found a mismatch between Lin's model and data, a confound between sequencing batch and species that undermined the conclusions\,\cite {gilad.mizrahi-man:2015:reanalysis}. This story is sadly not an exception. Data analysis is grappling with a reproducibility crisis.  Many causes underlie the crisis, but software bugs play a major role\,\cite {soergel:2015:rampant, joppa.ea:2013:troubling, robertson.ea:2025:scientists}.  A standard answer from the software engineering research field to such problems is to understand and advance software quality assurance techniques, chiefly testing, for the programs in the domain.

A number of papers address bugs in probabilistic programs and probabilistic inference systems\,\cite {storm:dutta2019,probfuzz:dutta2018}, however, to our knowledge, none attacks the \emph {oracle problem}\,\cite {barr.ea:2015:oracle} or, more generally, formalizes properties to automate the testing of probabilistic models. Correctness is hard to specify here. It is difficult to discern whether poor convergence or poor fit arise because the problem is hard, the model is broken, or wrong assumptions were made. It is thus hard to understand if the numbers produced by inference carry ground-breaking discoveries or manifest bugs.  This calls for using partial proxy properties as tests. Yet it is unclear which such properties hold for data analysis models in general and which only for specific model classes. Answering this would enable the systematic reuse of tests and prevent checks where they do not apply. Finally, there is a gap between the rich but largely manual and informal practices for checking models used in applied statistics and the automated, systematic testing practices of software engineering.

\begin {example*}

\Cref{fig:example} presents the classic dilution assay model of Fisher.  The goal is to estimate the concentration $\rho$ of organisms in a liquid, originally protozoa in soil samples~\cite {fisher:1922}.  The idea is to extract $N$ small identically sized samples of the liquid, incubate them, and see which ones show growth.  The groups of $N=4$ samples are shown as circular Petri dishes in the figure.  Each sample that grows contains at least one organism.  If the liquid is well mixed, the number of organisms in each sample is Poisson-distributed, so the probability of no growth is the probability of value zero in the Poisson distribution.  Unfortunately, if the liquid is rich, all samples grow, and we learn nothing of concentration.  So to decide how big $N$ should be, we need to have a pretty good idea of what the concentration $\rho$ is (a chicken-and-egg problem), and given that a suitable $N$ is exponential in $\rho$, it might not be practical for high concentrations.  To work around this issue, we fix $N$, but run the experiment $K$ times, diluting the liquid 1:1 with clean water each time.  This eventually produces both sterile and growing samples, and does so exponentially fast, thanks to halving $\rho$ at each step.  All $N \! \times \! K$ observations then contribute to estimating~$\rho$.  We use a binomial variable to relate the counts of growing samples to probability $\pi_i$ of growth at step $i$.  The complement, $1-\pi_i$, is the Poisson probability of a sample containing no organism.  These probabilities, transformed through a complementary-log-log link, give the linear predictor in the figure.  The model is an instance of a GLM, a rich class of models widely used in science.  Other common GLM examples are linear regression, logistic regression (classification), and count models.

The bottom part of \cref {fig:example} shows the model implemented in PyMC, a popular probabilistic programming DSL embedded in Python\,\cite {abril-pla.ea:2023:pymc}.  The program is short but dense.  Real-world models, say in genetics or materials science, can span hundreds of lines, and their correctness properties are non-trivial.  Unfortunately, ordinary errors in such programs can produce hard-to-detect symptoms, like a quietly wrong posterior with a plausible mean, a believable histogram, and no crash.  Both the cognitive challenges and the lack of clear functional failure signal call for a stringent quality assurance process for these models.

\end {example*}

\noindent
The field of applied statistics has long recognized this, producing a range of advice on what to look for.  For instance, for the example model it would be useful to check that the uninformative prior for $\rho$ indeed translates to a broad prior for $\pi$. We call this property \emph{Prior behavior on outcome scale} later in the paper.  It is particularly important for non-trivial link functions like the one in our example (as it turns out, this model hides a small bug here).  One should also check that inference indeed works in the model, so that data fitting actually narrows the prior into the posterior (\emph{Information content of the likelihood}). A model where this is not happening is likely wrongly wired, representing the wrong dependencies.  Furthermore, a common practice is to generate synthetic data with known parameter values ($\rho$ in this case), and check whether the model can recover them (\emph{Parameter reconstruction}).  Unfortunately, the classic statistical modeling texts do not operationalize such advice as executable oracles~\cite{gelman.ea:2026:workflow}, and do not yet systematically introduce the practice of automatic regression testing into building the models.  Tests are manual, often visual (compare plots), and rely on human judgment (``the values seem to be close enough to each other'').  The community has however called for increasing the automation\,\cite{steinruecken.ea:2019:automatic}.
\looseness -1

In this work, we study testable properties of probabilistic models corresponding to the broad class of generalized linear models (GLM) \cite{bda,regression-and-stories, mccullagh.nelder:1989}.  We identify correctness properties by analyzing \emph {Statistical Rethinking}~\cite[hereafter SR]{mcelreath:2020:statistical}, likely the most widely used comprehensive text on Bayesian modeling in the education of data analysts across many scientific domains.  We select executable oracles, implement them in Python and test them against a newly assembled collection of 59 PyMC models. This allows us to study which of the guidelines included in SR are automatable as properties, to what extent we can identify models to which they apply, whether they are falsifiable (we simulate regression testing with positive and negative model examples for each test), whether the tests are flaky in practice, and whether the faults they guard are captured in the fault model of the mainstream mutation tool for Python \emph{Cosmic Ray}.  Our contributions are:
\begin {enumerate}

  \item A comprehensive collection and automation of tests for probabilistic programs extracted from SR, a widely-used text on creating Bayesian data analysis models.

  \item An investigation of the suitability of the modeling guidelines  as a source for automatic tests (what guidelines are automatable, how efficiently, how flaky the resulting tests are, whether the standard mutation captures the bugs tracked by these properties)

\end {enumerate}

\noindent
Overall, we identify 17 specification properties applicable across all main aspects of the probabilistic models (data, prior, likelihood, posterior, parameters, model structure). Most of the properties are broadly applicable to many models or model classes. We show that all properties are relevant and testable on realistic models, using our designed oracles, and that these oracles are robust, in the sense that most of the tests are not flaky.  This is despite the inherent non-determinism in both models and tests.  We gather evidence that flakiness in this domain depends more on the complexity of the model-under-test than on the tested property. Finally, we discover a significant mismatch between the fault models induced by our probabilistic properties and standard mutation testing, calling for dedicated mutation operators for this domain.

Our method engages deeply with the domain knowledge of Bayesian data analysis and probabilistic programming as practiced in the field, using a textbook synthesis as a proxy.  We have two audiences. First, we want to demonstrate to the Bayesian Data Analysis community, that it is possible and beneficial to automate a large part of their procedures.  Second, we want to attract the software testing community to the properties and model requirements that data analysts work with, in order to ground the software engineering research for probabilistic programs in statistical modeling practice.
\looseness -1

The paper has two main parts. We first describe the elicitation process  and the resulting set of properties (\cref {sec:qualitative_analysis,tab:properties_summary_table}).  We then study these properties from the testing perspective (\cref {sec:experiments}).  The remaining sections cover threats to validity, related work, data availability, and conclusions (\cref {sec:threats_to_validity,sec:related,sec:data_availability,sec:conclusion}).
\looseness +1

\section {Synthesis of the Probabilistic Properties Dataset}%
\label {sec:qualitative_analysis}

\subsection {The Source}

We extract correctness properties of probabilistic models by reading and analyzing the entire text of \emph{Statistical Rethinking}\,\cite{mcelreath:2020:statistical}.  SR is a comprehensive teaching book targeting practitioners of Bayesian Data Analysis.  It reaches broad audiences as it is used for training researchers across fields. While precise sales data are not available, the online course based on the book accumulates tens of thousands of views.\footnote{e.g.\,\url{https://youtu.be/watch?v=ztbYkBPDOgU} or \url{https://youtu.be/watch?v=cclUd_HoRlo}}  SR's pedagogy is centered around modeling, model justification, diagnostics, comparison and selection, consistently following a fail-debug-improve cycle for probabilistic models.  Problem-by-problem, it presents common quality assurance advice that is built on many years of research, practice, and teaching.  The book covers a wide spectrum of probabilistic models, spanning from general models such as linear regression, to more specialized ones, such as mixtures and hierarchical models. Its examples, presented in R, have been translated by the community to several probabilistic programming languages, including PyMC and Pyro. Hence, the properties it prescribes are not tied to a single probabilistic programming language. All in all, SR is a prime source of language-agnostic quality assurance procedures that analysts in many fields are trained to apply (manually) today, to many classes of models. This makes it a rich and informative source for our objectives.
\looseness -1

\subsection {Analysis and Implementation}

We follow a two-phase design.  In the first phase, we identify correctness properties for model design.  In the second phase, we implement oracles to make the properties automatically testable, and to study how they behave as tests.

We begin by identifying procedures that SR recommends to assess whether a model or its inference behave as expected---we call these operations checks. For each check, we record its location in the book, the type of model it is applied to, the quantity computed, the reference value it is compared against, and how the outcome is judged.   We exclude checks for which the oracle is unclear, or impossible to implement without performing extensive property-specific research (to the best of our knowledge).  Chiefly, the checks that rely on subjective judgment that cannot be turned into an automated decision procedure are excluded.

The process is iterative.  We cluster and refine the properties of interest as we proceed, by comparing each identified check with the prior ones and adjusting them (for instance merging). Once the list is stable, we iterate again over the final collection to ensure consistency. \looseness -1

Three of the authors jointly analyze the SR text and identify the properties.  This improves coverage.  The authors have mixed expertise in statistics and data science, formal methods, and software engineering, so each reader tends to emphasize different properties. The resulting properties are then reviewed among all coauthors. Each disagreement over a property definition, its characteristics, or oracle implementation, is discussed until a consensus is reached.

SR typically judges the outcome of a check by visual or informal inspection, a human oracle.  To turn checks into tests, we implement the test oracles in PyMC and apply them to both correct and incorrect models, simulating regression scenarios.  Many of our tests follow the Bayesian paradigm of the models under test, but some are pragmatically frequentist.  Note that tests do not need to follow the same paradigm as the models under test, as long as they are able to identify problems. One can test non-point estimates using  point-estimates. In fact, a pass/fail verdict is a point estimate.  In properties that use Bayesian inference (prior-predictive, posterior, posterior-predictive), we choose Markov Chain Monte Carlo (MCMC) to infer samples \cite{bda}.  More details on the implementation and how it is used in further experiments is provided in \cref{sec:experiments} and in \cref{tab:properties_summary_table}.
\looseness -1

For each property we provide a short name, a definition, the main oracle implementation concept, the targeted model aspect (for instance the prior or the posterior), to what types of models it pertains (or whether it applies generally), and whether it is theoretically flaky (whether it has a positive probability of flipping the verdict without code change).

\subsection {Results}

After scouting, exclusion, and clustering the checks, we identified 17 properties   (\cref{tab:properties_summary_table} on \cpageref{tab:properties_summary_table}).

\subsubsection* {Oracle Implementations.}

We provide an explicit oracle for each property (\cref {tab:properties_summary_table}). Note that these oracles were not included in SR, which tends to defer to human expert judgment.  Supplying the oracles requires a shift of perspective. In data analysis, statistical hypothesis tests concern the parameters of a model given the data.  In software testing, they concern the model outputs across repeated runs\,\cite {arcuri.briand:2014:hitchhiker}.  For instance, \emph {Shrinkage level variance} requires running inference several times and reasoning about the variation across runs, a question one cannot pose from inside the model. Treating the model as an object of study rather than as a tool for studying data is routine in model comparison and selection, but data analysts do not routinely formulate pass/fail properties during these activities.  Identifying and implementing suitable statistical hypothesis-test oracles constituted a substantial part of this work.  Structuring them in a property pattern language\,\cite {dwyer.ea:1999:patterns} is a natural next step and would lower the barrier to entry for data analysts into automated testing.

\subsubsection* {Target.} Our properties can be grouped into:
\begin{enumerate*}[i)]

    \item distributional properties of prior, likelihood and posterior, such as unimodality or flatness;

    \item distributional properties of data, such as normality or zero-inflation,

    \item link function properties, such as type related properties,

    \item properties related to parameter dimensionality, such as size of parameter vectors,

    \item hierarchical properties of model structure, such as shrinkage properties, and

    \item structural properties for model efficiency, such as centered vs non-centered model design.

\end{enumerate*}
As shown in~\cref{tab:properties_summary_table}, 11 of the 17 properties check the posterior and the remaining properties address the likelihood and the prior.  Overall, this shows a good coverage of the Bayesian workflow, all the model aspects of concern.

\subsubsection* {Model Type (Generality)} Our properties include both general checks, applicable to different types of models, and specific checks for select model classes.  The generally applicable properties enforce universal statistical concepts that are relevant regardless of the model type (similar to say deadlock-freedom or lack of data races in concurrent programming). Our implementations of derived tests can be reused by data analysts using PyMC \emph {regardless of the model they work with}. The specific properties enforce requirements regarding a problem or a requirement for a class of models, for instance linear regression models. Six properties in our collection are general and apply to most model types and the rest are model-type-specific, most commonly limited to \emph{Count} and \emph{Categorical} models, which themselves are broad classes.

Some of the general properties target classic statistical concepts: \emph{Posterior unimodality} checks whether the posterior has a single mode, \emph{Directional bias of the posterior} looks at its symmetry, and \emph{Multicollinearity} flags strong correlation among the model's predictors. Others use a test procedure that does not depend on the model type. \emph{Information content of the likelihood} compares the prior and the posterior through their KL-divergence. \emph{Parameter reconstruction} verifies if the parameters used to simulate data can be recovered from it. \emph{Conditional independence} attempts to confirm if the model is consistent with its assumed causal structure. The last two properties rely on external input aside from the model itself: simulated data with known original parameters and previously assumed causal relations respectively.

Examples of model-type-specific properties include \emph {Data-distribution match}, which verifies whether the residuals of the model are normally distributed, an essential assumption of a linear model.  Several concern GLMs. A GLM uses a link function to connect a linear predictor, which ranges over all reals, to an outcome whose range is constrained, e.g.\ a probability in [0,1], or a non-negative count.  The example of \cref {sec:introduction} uses the complementary log-log function. \emph {Link function match} confirms whether the correct function is used for the data and the proposed likelihood. \emph {Prior behavior on outcome scale} validates the prior's distribution after passing through the link function, while \emph {Posterior-predictive distributional validity} applies the same check to the posterior predictive distribution.  There are properties that only apply to one family of GLMs: categorical models (e.g.\ Bernoulli and Multinomial) or count models (e.g.\ Poisson and Negative Binomial). \emph {Categorical variable validation} and \emph {K-1 model completeness} test whether the data and the model definition respect the correct categories (essentially a type check). \emph {Over-dispersion variance compliance} and \emph {Zero-inflation frequency} detect whether the posterior correctly follows the assumptions of a Poisson model; otherwise it should be replaced with a Negative Binomial model or a Zero-inflated model, respectively.
\looseness -1

The specific properties above concern the likelihood, and each model belongs to exactly one of these classes. The remaining specific properties concern the model's parameter structure instead, and can combine with any likelihood class: a model can be both Linear and Multivariate, or both Count and Hierarchical. Hierarchical models implement hyper-priors to model both populational and group-level variation in the model. \emph{Non-centered parameterization} validates whether the chosen parameterization (centered or non-centered) suits the data, and \emph{Shrinkage level variance} checks whether group-level estimates are pooled as expected. Multivariate models treat the outcome as a vector rather than a single value. \emph {Correlation matrix values} specifically tests whether the Correlation matrix, a prior parameter particular to these models, is correctly defined.

The structure of our collection, induced by target model type, provides useful signal for how a property pattern DSL for property-based testing should be structured, following the model types.

\subsubsection* {Theoretical Flakiness}

Several oracle implementations in our collection are based on statistical methods that are non-deterministic, as they use random sampling or their input varies over runs.   Randomized behavior might thus emerge not only from the models under test, but also from the oracle implementations. Consequently, some of our implementations can yield \emph{flaky tests}\,\cite{luo.ea:2014:fse}.  However, randomized procedures and models do not absolutely have to lead to observable of flakiness in practice (if the probability of non-deterministic verdict is sufficiently small). We thus distinguish two notions of test flakiness for probabilistic models.

We say that a test is \emph{theoretically flaky} if each test run has a non-zero probability of both fail and pass outcome, meaning repeated runs on its model can produce both results. We call a property theoretically flaky if applying it to a model can result in a theoretically flaky test. A test is \emph{practically flaky} if we observe both passing and failing outcomes when running it several times on the same model without any code or data change.  In our collection, 14 of the 17 properties are theoretically flaky.  We investigate their practical flakiness in \cref{sec:eval:rq3}.

\section {Analysis of the Probabilistic Properties as Tests}%
\label {sec:experiments}

\newcommand \rqone
  {Are our properties obtained from modeling guidelines testable on realistic models?}

\newcommand \rqtwo
 {Are the tests derived from modeling guidelines flaky in practice?}

\newcommand \rqthree
  {Is the fault model captured in the standard mutation testing tools consistent with the properties derived from modeling guidelines?}

We study the properties in our collection (\cref{tab:properties_summary_table}) from the perspective of automated testing. In particular, we ask the following research questions:
\begin{enumerate}[label=\textbf{RQ\arabic*}]
  \item \rqone\label{rq:1}
  \item \rqtwo\label{rq:2}
  \item \rqthree\label{rq:3}
\end{enumerate}

\noindent
Modeling advise in textbooks might be repeated as ``folk knowledge'' or it might be not automatable as tests.  The first research question verifies whether the guidelines from text books (as exemplified by SR) can be effectively tested on realistic models.  We answer it qualitatively by using our properties to test probabilistic code.  \ref{rq:2} investigates whether the flakiness of SR properties is only theoretical, or whether it manifests in practice.  We answer it quantitatively by running the test suites on realistic models multiple times. We address \ref{rq:3} in a quantitative experiment that investigates whether the properties can detect bugs introduced by mainstream mutation operators\,\cite{demillo.ea:1978:hints}. Mutation testing tools adopt a fault model that captures a class of errors in standard programming tasks.  We want to understand whether the standard mutation testing is applicable to probabilistic properties and models, so if it is able to violate probabilistic properties.  This is needed to establish evaluation procedures for research on testing methods for probabilistic models.  A negative result would mean that a new fault model for mutation is needed in this domain.

\begin{table} [t]

  \begin {tabular} {
    >{\small}p{3.5cm}
    >{\small}r
    >{\small}l
  }
      \textbf{Source} &
      \textbf{\# models} &
      \textbf{Type}
      \\
      \midrule
      Statistical Rethinking (SR) &
      29 &
      Academic (educational, realistic)
      \\

      PyMC docs &
      4 &
      Industrial (educational, realistic)
      \\

      PosteriorDB &
      26 &
      Academic (benchmarking, both real-world and realistic)
      \\
      \bottomrule

  \end {tabular}

  \smallskip

  \caption {Models for evaluation on each source}%
  \label {tab:model_sources}

  \vspace {-3ex plus 3pt minus 1pt}

\end{table}

\subsubsection* {Subject Probabilistic Models}%
\label {sec:subjects}

Since models and properties are paired in SR (each property is pedagogically exemplified on a matching model) it is important to find alternative cases to reduce bias.  We follow \cref{tab:properties_summary_table} performing a targeted search for models of the given types in order to identify examples to which the properties pertain.  Note that a random sampling would be unlikely to give a lot of interesting property--model matches, given how large the specification and model space is.  A random pairing of specifications and systems is unlikely to be successful in any domain.

We collect PyMC models from three sources: SR, the PyMC website,\footnote{\url{https://www.pymc.io/}} and the PosteriorDB\footnote{\url{https://github.com/stan-dev/posteriordb}} dataset.  PyMC is one of the most popular probabilistic programming frameworks.  It is expressive enough to handle all properties of \cref{tab:properties_summary_table}.  Even though SR presents the properties informally, the models are fully specified as programs in R using the \texttt{rethinking}  package.  We use independent translations of these models to PyMC.\footnote{\url{https://github.com/pymc-devs/pymc-resources/}}  PosteriorDB, a benchmark for Bayesian inference engines, contains equivalent models in multiple languages, including PyMC.  Most models are realistic, in the sense that they are sanitized versions of actual models from science and engineering.  PosteriorDB also contains some real-world, not sanitized, models. Its models are on average larger and more complex than of the other sources.  The resulting set extends beyond the scope of SR by reflecting more unusual applications (e.g., of hierarchical modeling) and architectures (e.g.,\,Bayesian non-parametrics, rare priors, likelihoods, and link functions). Our evaluation dataset comprises 59 models (\cref{tab:model_sources}).
\looseness -1

All models are designed by experts in probabilistic modeling (not ourselves),  and are supposed to be correct.  However, we do include (if available in a data source) or create (if not available) incorrect versions of some models to make sure that each property has at least one satisfying and at least one falsifying witness.  This emulates a regression test scenario, and shows that our oracle implementations are effective.  The negative examples represent common errors in model design, the positive examples represent the error resolutions. Overall the models can be used to evaluate whether our properties detect errors in more complex models than those of SR, and even models not originally in the scope of the book.

\subsubsection* {A Uniform Harness for Properties and Models (Implementation)}%
\label {sec:harness}

Since the relation between the model-dataset elements and the property-collection elements is many-to-many, we established a fixed Python API for models and properties to support matching models and properties for executions. All models can enter a standardized pipeline as they implement a shared base class. Each specialization specifies the observed data, model parameters, and sampling configuration. The base class also treats the prior, the likelihood, the posterior, and the posterior predictive as independent stages, each of which is compiled only when a property requests it. If a property only uses the prior for its evaluation, it never infers the posterior. Once a stage has been sampled, its results are cached, so when several tests check the same model they reuse the existing samples rather than compiling and sampling again.  We wired the properties with models using Pytest.\footnote{\url{https://pytest.org}} This makes experiments reproducible and efficient, minimizes the chance of code-cloning bugs, and allows us to to make targeted changes to a part of a model (some properties require counterfactual model interventions or access to individual model parts).
\looseness -1

\begin{table} [t]

\setlength \tabcolsep  {2pt}

\begin{tabularx} \linewidth {
  >{\small}X
  >{\small}R{15mm}
  >{\small}R{15mm}
  >{\small}R{15mm}
  >{\small}R{15mm}
}

\strut \newline \textbf {Property}
& \textbf {Model Type}
& \textbf {\#Passing witnesses}
& \textbf {\#Failure witnesses}
& \textbf {\#Nonflaky Passing}
\\
\midrule
\emph{Posterior unimodality}                        & any         & 8         & 7 & \textbf 6 \\
\emph{Directional bias of the posterior}            & any         & 8         & 3 & \textbf 7 \\
\emph{Multicollinearity}                            & any         & 10        & 11& 10\\
\emph{Data-distribution match}                      & linear      & 8         & 7 & 8 \\
\emph{Information content of the likelihood}        & any         & 9         & 5 & \textbf 8 \\
\emph{Categorical variable validation}              & cat.        & 3         & 1 & 3 \\
\emph{Prior behavior on outcome scale}              & count, cat. & 1         & 1 & 1 \\
\emph{K-1 model completeness}                       & cat.        & 3         & 1 & 3 \\
\emph{Posterior predictive distributional validity} & count, cat. & 3         & 1 & 3 \\
\emph{Parameter reconstruction}                     & any         & 2         & 1 & 2 \\
\emph{Link-function match}                          & count, cat. & 5         & 4 & 5 \\
\emph{Over-dispersion variance compliance}          & count       & 4         & 2 & 4 \\
\emph{Zero-inflation frequency}                     & count       & 1         & 1 & 1 \\
\emph{Non-centered parameterization}                & hier.       & 1         & 2 & 1 \\
\emph{Shrinkage level variance }                    & hier.       & 3         & 1 & 3 \\
\emph{Correlation matrix values}                    & mult.       & 1         & 1 & 1 \\
\emph{Conditional independence}                     & any         & 2         & 2 & 2 \\
\bottomrule
\end{tabularx}

\medskip

\caption{The number of models that display the respecting property (testability witnesses for \ref{rq:1} are model-property pairs). The last columns shows the number of practically non-flaky tests (\ref{rq:2}).}%
\label{tab:passing_models}

\vspace {-3ex plus 3pt minus 1pt}

\end{table}

\subsection {\ref{rq:1}: \rqone} \label{sec:eval:rq1}

We manually inspected our sources and selected models to which the properties from \cref {sec:qualitative_analysis} apply.
We consider a property applicable to a model when three conditions hold.
\begin{enumerate}
    \item The model belongs to the class the property targets. General properties satisfy this condition for every model.
    \item The model contains the components the property examines. For example, at least two predictors for \emph{Multicollinearity} or a categorical variable for \emph{Categorical variable validation}.
    \item The model includes inputs the property requires beyond itself, such as simulated data and known true parameters for \emph{Parameter reconstruction}.
\end{enumerate}

\noindent
A property is relevant and testable if it can be found applicable in at least one model of our external sources (non SR) and if we can demonstrate both a success and failure.  If a property (such as the general properties, marked \emph{any} in \cref{tab:properties_summary_table}) applies to many potential models, we do not test them exhaustively, but qualitatively select the external models that maximize diversity.

\paragraph{Results}
\Cref{tab:passing_models} summarizes the number of models for which we demonstrated testability, split into positive and negative test subjects. Note that we successfully demonstrate relevance and testability for each  property of \cref{tab:properties_summary_table} (we found at least one positive and at least one negative case for each of them).  As expected, general properties, with the previously discussed exceptions of \emph{Parameter reconstruction} and \emph{Conditional independence}, have a higher amount of models which range from simple linear regressions to more complex programs like Mixtures and State Space Models. Properties of Count models, even though not general, also apply to rather large families of models, with multiple likelihoods to test, which is reflected in the table.
We thus answer \ref{rq:1} positively.
\looseness -1

\begin{rqconclusion}
The properties derived from modeling guidelines, including our selected computable oracles,  are relevant and testable on realistic models, with varying levels of complexity.  The collected properties can thus be used as specifications in real projects, and can enrich testing frameworks and languages for probabilistic programs.
\end{rqconclusion}

\subsection {\ref {rq:2}: \rqtwo}%
\label {sec:eval:rq2}

\newcommand \klprop
  {\emph {Information content of the likelihood}}

\newcommand \uniprop
  {\emph {Posterior unimodality}}

\newcommand \skewprop
  {\emph {Directional bias of the posterior}}

\newcommand \checksymbol
  {\ding {51}}

\newcommand \crosssymbol
  {\ding {55}}

\begin {table} [t]

  \begin {tabularx} \linewidth {
    >{\small}X
    >{\small}R{15mm}
    >{\small}R{18mm}
    >{\small}R{15mm}
  }
    \textbf {Property} \( \rightarrow \) \newline
    \textbf {Model (Source)} \( \downarrow \) &
    \emph {Information content}  &
    \emph {Posterior unimodality}    &
    \emph {Directional bias}
    \\
    \midrule

    {\small GsrmLatentRegItr (PosteriorDB)} &
    \checksymbol &
    NA &
    NA
    \\

    {\small SeedsCentered (PosteriorDB)} &
    \checksymbol &
    \crosssymbol &
    NA
    \\

    {\small NormalMixtureK (PosteriorDB)} &
    \checksymbol &
    \crosssymbol &
    \crosssymbol
    \\

    {\small EightSchoolCentred (PosteriorDB)} &
    NA &
    \checksymbol &
    \crosssymbol
    \\

    {\small StateSpaceStochasticLevelStochasticSeasonal (PosteriorDB)} &
    NA &
    \checksymbol &
    \checksymbol
    \\
    \bottomrule

  \end{tabularx}

  \medskip

  \caption{Tests showing practical flakiness. The symbol \checksymbol\ means that the (property, model) is flaky in the experiment, \crosssymbol\ denotes no practical flakiness observed. NA: our collection did not include the specific pair.}%
  \label{tab:flaky_qualitative}

  \vspace {-3ex plus 3pt minus 1pt}

\end{table}

Several factors influence whether a theoretically flaky test (\cref{sec:qualitative_analysis}) manifests flakiness in practice: the tested property, its oracle choice, the probabilistic model under test, and the number of samples use to estimate distributions (the strength of statistical inference).  In this section, we evaluate practical flakiness for the properties of \cref{tab:properties_summary_table}.

We begin by constructing a set of tests. We say that a test is a pair of a model and an applicable property (choosing models and properties from the sets described above).  We use both passing and failing tests for the flakiness experiments (\cref{tab:passing_models}, 72 passing, 51 failing witnesses, 123 tests in total).

To ensure the accuracy of analysis, we exploit statistical bounds based on the coupon collector's problem~\cite{Christensen:2025:ProbTest}. We run each test 200 times. This number of executions is provided by that method and it ensures that, if a test does not show practical flakiness, then it is not practically flaky with probability at least 0.97. Thus, the probability of misclassifying a practically flaky test as non-flaky is below 0.03, and this holds with probability \( 0.999 \)~\cite{Christensen:2025:ProbTest}.

For properties that perform inference (most), we use the default parameter values of PyMC: 500 tuning steps, 1000 samples and 4 chains. In principle, increasing the sample size, should decrease flakiness, but we stick to the standard setting to avoid a more arbitrary choice.
\looseness -1

\begin{table} [t]

   \begin {minipage} [b] {0.568\textwidth}
    \begin {tabular} {
      >{\small}l
      >{\small}r
    }
    \textbf{Model} & \textbf{\llap{Fl}akiness \%} \\
      \midrule
      GsrmLatentRegItr & 50\% \\
      SeedsCentered & 33\% \\
      NormalMixtureK & 33\% \\
      EightSchoolCentred & 25\% \\
      StateSpaceStochasticLevelStochasticSeasonal & 50\% \\
      \bottomrule
    \end{tabular}
    \vspace{0pt}
  \end {minipage}%
  \hspace{1.8em}%
  \begin {minipage} [b] {0.347 \textwidth}
    \begin {tabular} {
      >{\small}l
      >{\small}r
    }
      \textbf {Property} & \textbf {\llap{Fl}akiness \%} \\
      \midrule
      \emph {Information content}   & 21\% \\
      \emph {Posterior unimodality} &  9\% \\
      \emph {Directional bias}      &  9\% \\
      \\
      \\
      \bottomrule
    \end{tabular}
    \vspace{0pt}
  \end{minipage}

  \medskip

  \caption{Fraction of flaky tests for models and for properties; 0\% for all other models and properties.}%
  \label{tab:flaky_percentage_marginals}

\end{table}

\paragraph{Results}
Only six out of the 123 tests ($<5\%$), pertaining to only three out of 17 properties, are flaky at runtime, despite most of the properties and tests being theoretically flaky.  This speaks for the robustness of our statistical oracles. \Cref{tab:flaky_qualitative} details the flaky tests and properties. \klprop{} shows flakiness for three models, two hierarchical categorical models and one mixture. \uniprop{} is flaky for two models, a hierarchical linear model and a state space model. \skewprop{} is flaky when testing the state space model. All the flaky tests in practice include models that come from PosteriorDB and have very complex parameterizations or likelihood functions. This indicates that avoiding flakiness is effective by using robust statistical procedures, but could be more difficult with larger or more complex models.

No single specific property is inherently prone to practical flakiness. The three properties with flaky tests are general properties that were applied to larger number of models, not displaying flakiness for many of them. \Cref{tab:flaky_percentage_marginals} (right) shows the percentage of tests showing flakiness, i.e., \(\nicefrac {\#\textit{flaky\_tests}_p}{\#\textit{tests}_p}\) for a property \(p\). \klprop\ shows the highest percentage of flaky tests, 21\%, i.e., 3 out of 14 tests show flakiness. In \uniprop\ and \skewprop\ the percentage of drops to 9\% (2 out of 21 and 1 out of 11 respectively). Properties themselves are not inherently flaky, and other factors explain the presence of flakiness.
\looseness -1

In our experiments, the model under test affects flakiness more than the property. For instance, GrsmLatentIgrModel was used for testing two properties and it shows a 50\% flakiness, with 100\% for theoretically flaky properties (the other property is theoretically not-flaky). This could be explained by the sampler configuration (the number of samples) not being sufficient for this complex model, or by the property tolerance threshold being set too high. Half (50\%) of the tests of StateSpaceStochasticLevelStochasticSeasonalModel are also flaky. This model was tested with three theoretically flaky properties and one deterministic test. However, only \uniprop{} and \skewprop{} show flakiness in practice for this model. The SeedsCenteredModel and NormalMixtureKModel were tested against three theoretically flaky properties and only one (\klprop{}) showed flakiness, i.e., 33\% flakiness rate. EightSchoolCentredModel was tested in four theoretically flaky properties, but only shown flaky for one (\uniprop). The other three properties, \klprop{}, \emph{Shrinkage level variance} and \skewprop{}, did not yield flaky tests. All in all, there are strong indications that the model affects flakiness more than a property, with complex models being more prone to flakiness.

Our answer to~\ref{rq:2} is thus that theoretical flakiness rarely appears in practice for the selected oracles and models, and most tests are not flaky.
\looseness -1

\begin{rqconclusion}
  Only 6 out of 123 tests were practically flaky, which indicates that statistical oracles can be robustly implemented, and flakiness associated with probabilistic programming might be possible to overcome in practice.
  Furthermore, the tested properties seem to have lower effect on the flakiness than the complexity of the models under test.
\end{rqconclusion}

\begin{figure} [t]

  \includegraphics [
    width = .49 \linewidth,
    valign = t,
    trim = 0mm 7mm 0mm 2.3mm,
    clip,
  ]{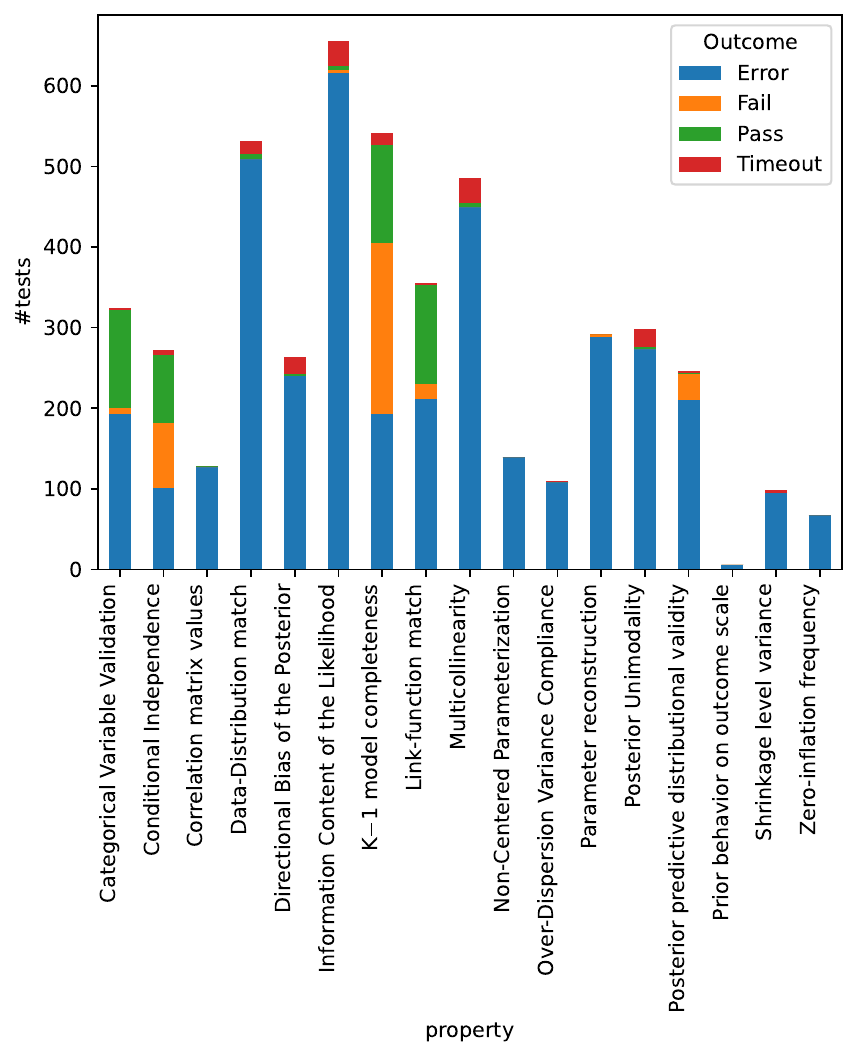}%
  \hfill%
  \includegraphics [
    width = .4817 \linewidth,
    valign = t,
    trim = 0mm 0mm 0mm 2.3mm,
    clip,
  ]{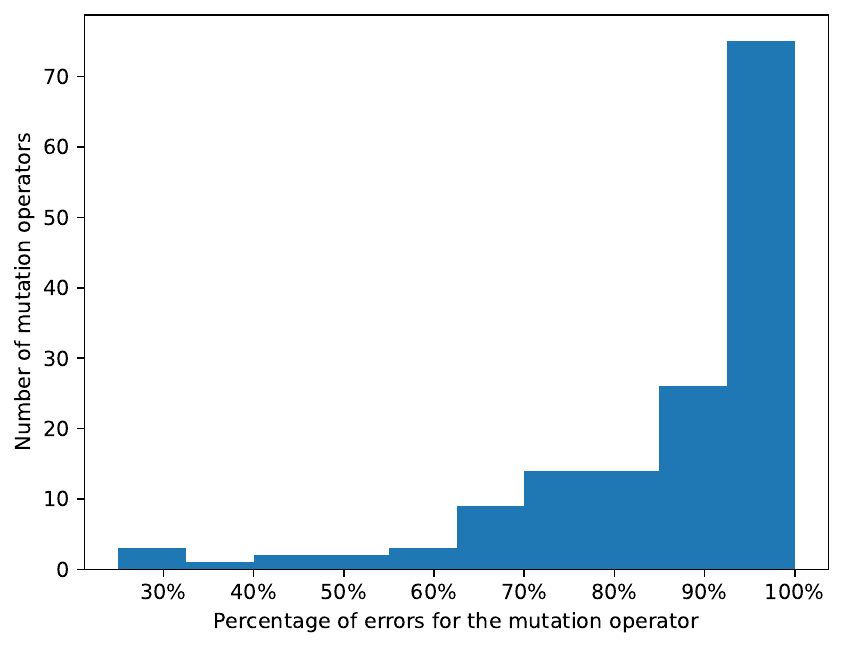}

  \vspace {-1ex plus 2pt minus 1pt}

  \caption {Left: Outcome for mutation test runs grouped by property.
  Right: Distribution of number of mutation operators with a given percentage of ``Error'' outcomes.}%
  \label{fig:prop_vs_test}

\end{figure}

\subsection {\ref{rq:3}: \rqthree}%
\label{sec:eval:rq3}

We use \emph{Cosmic Ray},\footnote{\url{https://cosmic-ray.readthedocs.io/}} a state-of-the-art general mutation testing tool that captures common coding errors in Python and integrates with Pytest. We use only the passing non-flaky tests for mutation testing; we disregard the text that exhibit practical flakiness in \cref{sec:eval:rq2}. This gives 68 tests  and ensures that  failures in the experiments stem from mutations (see the last column in~\cref{tab:passing_models}).

Cosmic Ray offers 161 different mutation operators. A large number of mutation operators replace binary arithmetic, bitwise and shift operations with each other. For instance, the operator \texttt {ReplaceBinaryOperator\_Mul\_Sub} replaces multiplication with substraction, or the operator \texttt {ReplaceBinaryOperator\_Div\_BitXor} replaces division with a bitwise XOR operation. Similarly, a group of mutations replace binary predicates such as equality or ordering relations (e.g., \texttt <).  Finally, there are mutations that replace numerical and Boolean constants (e.g., \texttt {Number\_Replacer}), include negations in Boolean predicates (\texttt {AddNot}) or change program statements (\texttt {ZeroIterationForLoop}). These mutation operators simulate common coding errors in regular programming tasks.

\begin{table}[t]

  \begin{tabular}{
    >{\small}l
    >{\small}r
    >{\small}r
  }
    \textbf{Mutation operator}              & \textbf{\% error} & \textbf{\#mutations} \\
    \midrule
    ReplaceComparisonOperator\_Gt\_Lt       & 25\%                 & 4           \\
    ReplaceComparisonOperator\_Gt\_LtE      & 25\%                 & 4           \\
    ReplaceComparisonOperator\_Gt\_GtE      & 25\%                 & 4           \\
    ReplaceComparisonOperator\_Eq\_Is       & 33\%                 & 3           \\
    AddNot                                  & 40\%                 & 10          \\
    ReplaceComparisonOperator\_Gt\_NotEq    & 40\%                 & 5           \\
    ReplaceComparisonOperator\_Eq\_GtE      & 50\%                 & 12          \\
    ReplaceComparisonOperator\_Eq\_Gt       & 54\%                 & 13          \\
    ReplaceComparisonOperator\_Eq\_Lt       & 56\%                 & 16          \\
    ReplaceBinaryOperator\_LShift\_Add      & 60\%                 & 5           \\
    ReplaceComparisonOperator\_Eq\_NotEq    & 62\%                 & 13          \\
    ReplaceComparisonOperator\_Eq\_LtE      & 63\%                 & 16          \\
    ReplaceBinaryOperator\_FloorDiv\_Mul    & 63\%                 & 8           \\
    ReplaceBinaryOperator\_FloorDiv\_BitAnd & 63\%                 & 8           \\
    ReplaceBinaryOperator\_Mul\_Sub         & 66\%                 & 73          \\
    ZeroIterationForLoop                    & 68\%                 & 6           \\
    ReplaceTrueWithFalse                    & 68\%                 & 3           \\
    ReplaceBinaryOperator\_Mul\_Mod         & 67\%                 & 76          \\
    ReplaceBinaryOperator\_Mul\_Add         & 68\%                 & 76          \\
    ReplaceBinaryOperator\_Mul\_BitAnd      & 70\%                 & 76          \\
    \bottomrule
  \end{tabular}

  \medskip

  \caption{Mutation operators least likely to cause errors and their counts}
  \label{tab:operator_error}

  \vspace {-3ex plus 3pt minus 1pt}

\end{table}

We instruct Cosmic Ray to apply all its mutation operators to each model (the properties remain intact). We obtain 3797 mutated models, and then execute the corresponding 4273 tests (some models have more than one property tested).

We set a timeout of 240 seconds per test run to determine whether a mutation introduces an excessive runtime overhead, and to make the experiment feasible.  This value is four times the average runtime of our tests, \(\approx60\) seconds with over 80\% of runs terminating below 120 seconds.  If a test run exceeds the timeout, it is counted as an error induced by mutation.

\paragraph {Results}

The left plot in \cref{fig:prop_vs_test} summarizes the distribution of outcomes for all tests, per property. Outcomes of type ``Error'' are the most prominent: runtime errors that appear before any test assertion is executed. In Python, runtime errors include issues such as division by zero or overflows, but also type errors.  This means that most mutations invalidate the model structure, and are \emph {unable to capture subtle modeling errors represented by the properties in our collection}.

\begin{figure} [t]

    \includegraphics [
      width = .5 \linewidth,
      clip,
      trim = 0mm 7mm 0mm 0mm
    ]{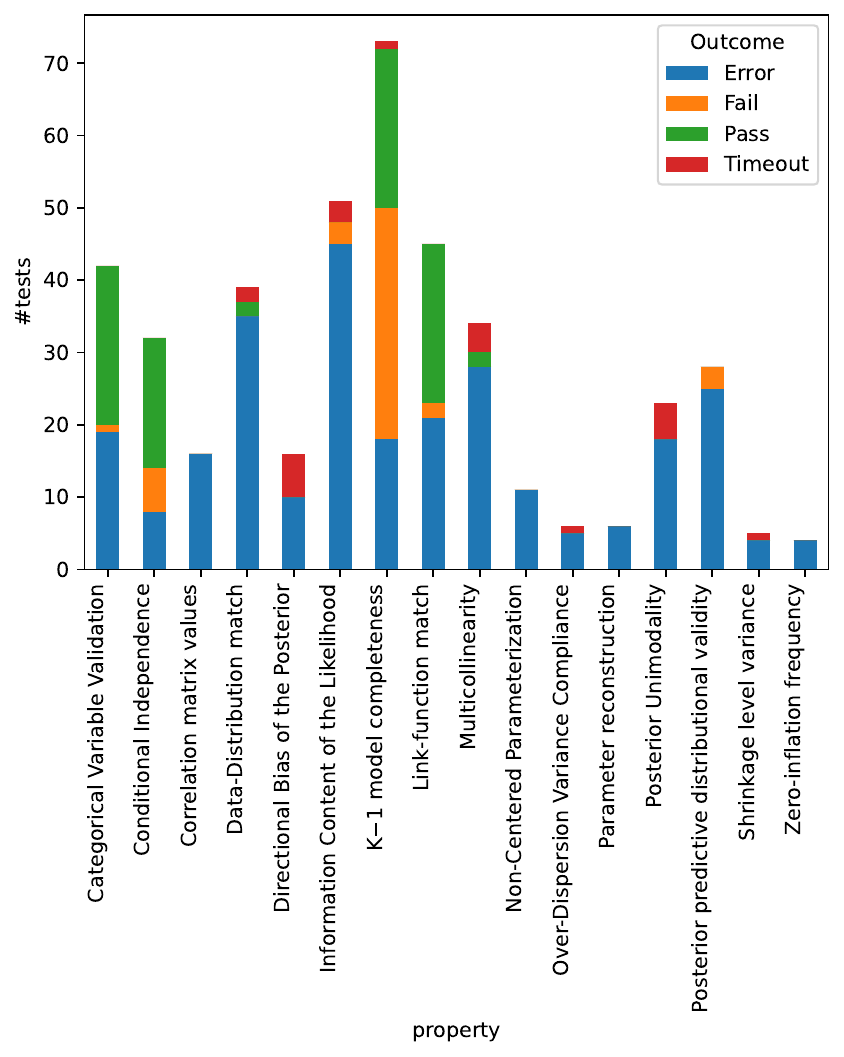}

    \vspace {-.8em plus 3pt minus 1pt}

    \caption{Outcome of mutation tests for operators with at most 70\% frequency of errors.}%
    \label{fig:prop_vs_test_70}

\end{figure}

In \cref{fig:prop_vs_test} (left), ``Pass'' labels tests where mutants survive (no runtime error, all assertions pass, termination before timeout).  For these cases, we additionally check KL-divergence between inferred prior, posterior and posterior predictive distributions of mutated and non-mutated models, or perform a manual check to determine if the property was violated by the mutation (for non-inference properties).  This is to understand whether mutants survive because the tests are weak or because the mutation is irrelevant. In none of the surviving cases, the mutation had any significant impact on the produced distributions.  \emph {The mutations and the specification properties are misaligned}---the properties from statistical guidelines concern distributions, while mutations are unable to induce distributional changes in many models.

``Fail'' counts tests that run to completion and trigger a violation of a test assertion (the mutant is killed).  Killing mutants is a positive verdict in mutation testing. It shows that the test harness works. Most of observed failures are for \textit{K-1 model completeness}, which tests a vector size.  Several operators, such as \texttt{Number\_Replacer}, change the integer used to set the size of a vector in the models.  More interestingly, most of the remaining properties kill mutants rarely or never.   This is another manifestation of a \emph{misalignment between the probabilistic specifications and general mutations}; this time showing that probabilistic properties are unable to catch most standard programming errors.
\looseness -1

Finally, timeouts constitute a small proportion of the test outcomes. These likely arise from mutants that produce an excessive increment in sampling time.

The rightmost plot in \cref{fig:prop_vs_test} shows the count of mutation operators that produce a certain percentage of ``Error'' outcomes. For 141 mutation operators the number of ``Error'' outcomes is over 70\%. \Cref {tab:operator_error} lists the remaining 20 operators that produce ``Error'' rarely (more likely to succeed in mutation testing). Most of these manipulate Boolean predicates. This is expected, as this operators do not change the type of an expression. Even after focusing on these operators the outcome distribution is still dominated by ``Error'' (cf.~\cref {fig:prop_vs_test_70}). The second most likely outcome type in this group is ``Pass''.  The prevalence of ``Error'' outcomes and insufficiently altered models further corroborates that the fault model embodied by these mutations is ineffective for testing our properties. Even a better aligned subpopulation of the operators is not effective.

Specific operators are more likely to yield ``Fail'' outcomes (\cref{tab:mut_fail}). They show the closest fault overlap with probabilistic properties. The \texttt {NumberReplacer} operator is by far the most likely to cause an assertion fail, with 133 detected failures. Admittedly though, this is the operator most frequently applied by Cosmic Ray in our experiments (1526 times), and its effectiveness in detecting failures is unsuprising, as many errors in probabilistic models arise from selecting incorrect parameter values (e.g., mean or standard deviation in prior distributions). These hints that mutations that affect parameter values can be useful for mutation testing of probabilistic properties.

Our answer to \ref{rq:3} is negative. As most of the 161 mutation operators result in runtime errors, most of the surviving mutants do not affect inferred distributions, and most of the properties never fail a mutant, we conclude that there is a significant misalignment between general programming and probabilistic programming fault models.

\begin{rqconclusion}
  Standard mutation testing mainly produces runtime errors when applied to probabilistic programs. Mutants that do not produce runtime errors rarely affect defined distributions, so they cannot exercise probabilistic properties.  Most probabilistic properties never fail on a mutant.  The domain of probabilistic models needs its own fault models, and mutation testing for probabilistic models should include \emph {both} standard operators (to detect standard bugs) and new probability-specific operators (to detect distributional issues). Mutation operators that focus on altering parameter values are the most effective, so they should be included in mutation testing tools.
  \looseness -1
\end{rqconclusion}

% \usepackage[normalem]{ulem}

\begin{table} [t]

\begin {tabular} {
  >{\small}l
  >{\small}r
}

\textbf{Mutation operator}           & \textbf{\#Failures} \\
\midrule
NumberReplacer                       & 133                 \\
ReplaceBinaryOperator\_Mul\_RShift   & 11                  \\
ReplaceBinaryOperator\_Add\_BitOr    & 9                   \\
ReplaceBinaryOperator\_Mul\_BitXor   & 8                   \\
ReplaceBinaryOperator\_Mul\_LShift   & 7                   \\
ReplaceBinaryOperator\_Mul\_BitAnd   & 7                   \\
ReplaceBinaryOperator\_Sub\_RShift   & 7                   \\
ReplaceBinaryOperator\_Sub\_BitAnd   & 7                   \\
ReplaceBinaryOperator\_Mul\_BitOr    & 6                   \\
ReplaceBinaryOperator\_Sub\_Mul      & 6                   \\
ReplaceBinaryOperator\_Add\_RShift   & 6                   \\
ReplaceBinaryOperator\_Sub\_Add      & 6                   \\
AddNot                               & 6                   \\
ReplaceBinaryOperator\_Add\_BitXor   & 5                   \\
ReplaceBinaryOperator\_Mul\_FloorDiv & 4                   \\
ReplaceBinaryOperator\_Sub\_BitXor   & 4                   \\
ReplaceBinaryOperator\_Add\_BitAnd   & 4                   \\
ReplaceBinaryOperator\_Add\_LShift   & 4                   \\
ReplaceBinaryOperator\_Sub\_Div      & 4                   \\
ReplaceBinaryOperator\_Sub\_LShift   & 4                   \\
ReplaceBinaryOperator\_Sub\_BitOr    & 4                   \\
% Others                               & 69                  \\
\bottomrule
\end{tabular}

\medskip

\caption{Mutations operators that created failures (interacted with probabilistic properties)}%
\label{tab:mut_fail}

\vspace {-3ex plus 3pt minus 1pt}

\end{table}

\section {Threats to Validity and Limitations}%
\label{sec:threats_to_validity}

\subsubsection*{External Validity / Transferability}

Our property collection comes from a single source, but SR is one of the most widely used comprehensive textbooks on Bayesian data analysis, covering a large variety of probabilistic models and properties.  We use 59 models from three sources, a number that is substantial given the manual refactoring and test implementation effort, but cannot represent all models in the ecosystem. Other models and properties could show different challenges in automation, with different flakiness and mutation behavior.  More important than completeness in this respect are the research process and the insights obtained: what kinds of properties can be found by studying literature, whether they are testable (\ref{rq:1}), whether they tend to be flaky (\ref{rq:2}), whether they track known or new faults (\ref{rq:3}).
\looseness -1

The mutation experiments in \cref{sec:eval:rq3} rely on a single tool, Cosmic Ray. Its mutation operators are representative of core operator categories used in many tools, but we have not verified whether other tools and languages would give the same results.

We used targeted search (\cref{sec:experiments}) rather than comprehensive pairing, so the tested property–model pairs reflect our judgment about which properties pertain to which models. The flakiness and mutation rates in \ref{rq:2} and \ref{rq:3} therefore describe this collection rather than estimating rates over the space of all plausible pairs. We note, though, that pairs were fixed before any tests were run, so the selection could not be influenced by flakiness or mutation outcomes.

\subsubsection* {Internal Validity / Credibility}

The properties of \cref {tab:properties_summary_table} may not include all automatable properties from SR, due to human interpretation of the text and the applied filtering.  We used best judgment to select automatable guidelines, cluster, and generalize them into a small set.  Three authors participated in the process (independent coding of each chapter in pairs and consensus meetings), relying on our expertise in Bayesian statistics, software testing, formal methods and domain analysis.
\looseness -1

The subject models are implemented in PyMC, although SR models and some of the PosteriorDB models were originally implemented in R.  We rely on established community translations of these models to PyMC, endorsed by the author of SR and created independently of this work. These translations are maintained in open source projects and widely used for research and teaching.  We refactored these models to follow our unified harness API (\cref{sec:experiments}), which may have introduced errors.  The manual inspection of failing tests and surviving mutants reduces the possibility of errors---we did not rely on test reports blindly.

\subsubsection* {Construct Validity}

Our oracles are not necessarily sound and complete representations of the modeling guidelines of SR.   First, the modeling guidelines are often informal and depend on expert judgment.  Second, we used approximations of the guidelines when operationalizing them. For example, the property \emph {Prior behavior on outcome scale} requires inferring the outcome scale distribution and checking it for flatness.  This can be done in many ways.  We chose to bin the distribution and lower-bound the probability mass in each bin.  This operationalization works for our cases, but it does not work for all models to which the guideline is applicable.  It would produce a spurious ``pass'' for a distribution with high-frequency variation within bins. Note that using partial or approximate specifications is a common pragmatic practice in testing.

The classification of properties as theoretically flaky depends on the oracle choice. There may exist non-flaky or less flaky oracles for some of the properties that we categorized as theoretically flaky.
\looseness -1

Using the default configuration to generate samples for tests may have increased the reported practical flakiness, since more draws would reduce Monte Carlo error.  However, the low observed practical flakiness (only 6 tests out of 123) bounds the effect of this decision.

\section {Related Work}\label{sec:related}

Prominent statistics authors\,\cite{gelman.ea:2026:workflow, steinruecken.ea:2019:automatic} recognize that \emph{developing a statistical model [...] is a form of software development} \cite[Chpt.\,15]{gelman.ea:2026:workflow}, with all the complexities and consequences that call for stringent design and quality-assurance practices.  Unfortunately, the sources they recommend are (largely) not based on systematic research-based knowledge.  With this paper, we want to contribute to systematization of the field of testing probabilistic models, by automating common advice, understanding the flakiness of the derived tests, and matching it against the fault models of a common mutation testing tool.  Our long term goal is to provide evidence-based domain-specific testing tools.
\looseness -1

\paragraph {Correctness Properties for Probabilistic Programs}

Formal methods research provides a host of specification languages, also for probabilistic systems.  Specification languages used in statistical model checking\,\cite {agha.palmskog:2018:survey,younes.simmons:2002:cav}, such as probabilistic process algebra~\cite {hillston:1996:pepa-book} and probabilistic temporal logics~\cite{hansson.jonsson:1994:pctl} cannot express the quantitative properties of interest in data analysis, where we handle complex multivariate continuous and discrete distributions with many data observations. They have been designed for specification of probabilistic properties of stateful control systems, not data analytics.  In contrast, our properties target data analysis models directly, so models that are multivariate and mix continuous and discrete dimensions following many classes of distributions.
\looseness -1

Grunske catalogued probabilistic properties as patterns\,\cite {grunske:2008:specification} in the sense of Dwyer et al.\,\cite{dwyer.ea:1999:patterns}. He assembled ProProST, a dataset of temporal properties of stateful probabilistic models (CTMCs, DTMCs) and the corresponding temporal logics (CSL and PCTL) as seen embedded control systems, web applications, and network protocols.  In contrast, we collect the requirements from a text (SR) that encodes years of established practice and hundreds of research papers in data modeling and analysis. In particular, our properties cannot be represented using ProProST's patterns, as the data analysis models and their properties differ wildly from those in probabilistic control.
\looseness -1

The PosteriorDB project\,\cite{magnusson.ea:2023:posteriordb} collects probabilistic data analysis models, which we use extensively in \cref{sec:experiments}. It was created to support development of quality diagnostics for statistical models, including testing implementations of Bayesian inference algorithms against each other. However, PosteriorDB itself does not contain any correctness properties, tests, or specifications beyond reference posteriors (which are oracles for inference algorithms, not for models).  Our collection thus complements theirs by specifying properties that the models themselves should satisfy.

\paragraph {Flakiness}

Flakiness often stems from concurrency and inter-test dependencies~\cite{luo.ea:2014:fse}, but flakiness induced by randomization is increasingly reported~\cite{dutta.ea:2020:issta}. For example, about 60\% of flaky tests in inference engines are caused by algorithmic non-determinism~\cite{dutta.ea:2020:issta}.  Several tools and studies \cite{dutta.ea:2020:issta,dutta.ea:2021:fse,dutta.ea:2021:issta} already consider identifying flakiness in probabilistic programs and reducing it.  These works take the departure in a collection of flaky tests, whereas we begin more broadly, with specification properties, and then learn that only few of the specifications are flaky. For example, Dutta and coauthors study flakiness in probabilistic programs, mostly deep-learning-based, using Pyro, PyMC, TensorFlow probability, and PyTorch\,\cite{dutta.ea:2020:issta} through the lens of 75 historical (practically) flaky tests.  Their conclusions are in agreement with ours: the most common cause for flaky tests is randomness in the considered systems.  They present FLASH, a tool for detection of flakiness.  Unfortunately, the tool, developed with PyMC3, is not maintained, and it no longer works with PyMC (we use v5), so we could not evaluate whether it is effective against our properties. Our work differs by having different objectives: we are interested in systematizing knowledge about correctness and testing, in order to work around the oracle problem, while we see flakiness as a relevant dimension of interest.  Furthermore, our subject systems are primarily analytical scientific models, not deep learning systems.
\looseness -1

\paragraph {Mutation Testing}

Mutation testing has been extended to a Bayesian setting for neural networks~\cite{humbatova.ea:2021:deepcrime,tambon.ea:2023:ist,jahangirova.tonella:2020:icst}, but these extensions offer no mutation operators for probabilistic faults (e.g., edits simulating misspecification of a distribution). A core difficulty is that a mutation's effect is confounded by the program's inherent randomness.   Recent work has begun to address this by using statistical tests to distinguish a mutant's effect from random noise~\cite{jahangirova.tonella:2020:icst,humbatova.ea:2021:deepcrime,tambon.ea:2023:ist}, but  only for a single scalar accuracy objective of a neural network, not for a test suite of many different distributional properties.   The presented properties \cref{tab:properties_summary_table} constitute the first step towards understanding what mutations would represent with common modeling problems.  Our experiments do demonstrate that the standard mutation operators are not sufficient.

\section {Conclusion}\label{sec:conclusion}

Our goal was to explore the use of modeling guidelines found in Bayesian data analysis literature for automated model testing. We demonstrate how to effectively elicit and automate correctness properties for probabilistic data analysis models.  Our collection contains 17 broadly applicable probabilistic properties, designed and implemented in Python and evaluated on 59 realistic models from three repositories.  We show that probabilistic tests, if setup as robust statistical procedures, do not need to be flaky; albeit achieving non-flakiness for more complex posteriors  is challenging (four PosteriorDB cases). We believe that we are the first to evaluate the effectiveness of mutation operators designed for regular coding tasks to assess specifications of probabilistic models. We point out deficiencies of standard mutation testing methods when applied in this field. The fault models of standard mutation operators and the identified properties are not compatible.  Most mutation operators introduce runtime errors that prevent tests from evaluating the assertions that capture the properties. Moreover, many operators alter the probabilistic model insignificantly, so they do not interact with probabilistic properties. This calls for the definition  of new mutation operators for this domain, and a suggestion to consider operators that affect parameter values.

In the future, we intend to exploit the results of this domain analysis to create a property language for a testing tool for probabilistic models. Furthermore, we are working on an adequate mutation testing methodology for probabilistic models.

\section {Data Availability}%
\label{sec:data_availability}

SR (the source) is available from the publisher and academic libraries. The replication package including all other material (subject models under test, properties, oracle implementations, analysis data, and scripts) is stored in a git repository that will be made open and archived on Zenodo upon acceptance.  An anonymized pre-release version is available for the discretion of the reviewers.\footnote{\url{https://anonymous.4open.science/r/protest_tests-0EF2/README.md}}

\bigskip

\setlength \tabcolsep  {2pt}
\setlength \LTpre      {0pt}
\setlength \LTpost     {0pt}
\setlength \LTcapwidth \linewidth
\setlength\LTleft{0pt}\setlength\LTright{\fill}

\begin{xltabular} \linewidth {
  @{}
  >{\footnotesize\it}P{.12 \linewidth}
  >{\footnotesize}P{.28 \linewidth}
  >{\footnotesize}P{.33 \linewidth}
  @{\hspace{2.5pt}}
  >{\footnotesize}P{.08 \linewidth}
  >{\footnotesize}P{.07 \linewidth}
  @{}
  >{\footnotesize}P{.08 \linewidth}
  @{}
}

\caption{Properties extracted from SR modeling guidelines, after filtering for implementable oracles (cat.=categorical, hier.=hierarchical, mult.=multivariate)}
\label{tab:properties_summary_table}
\\[-6.0mm]

  \textnormal{\textbf {\hspace{2cm} Name}}
  & \textbf {\hspace{4cm} Definition}
  & \textbf {\hspace{6cm} Oracle implementation}
  & \textbf {\hspace{1cm} Target}
  & \textbf {Model Type}
  & \textbf {Theoret.\ Flakine\rlap{ss}}
\\
\midrule
\endfirsthead

\multicolumn 6 {@{}l}
  {\small\textsf{Table \ref{tab:properties_summary_table}. (continued)}}
\\[-3mm]

\hiderowcolors
  \textnormal {\textbf {\hspace{2cm} Name}}
  & \textbf {\hspace{4cm} Definition}
  & \textbf {\hspace{4cm} Oracle implementation}
  & \textbf {\hspace{1cm} Target}
  & \textbf {Model Type}
  & \textbf {Theoret. Flakine\rlap{ss}}
\\
\midrule
\showrowcolors
\endhead

\endfoot

\endlastfoot

Posterior Unimodality &
The posterior has one mode &
Hartigan's dip test on the posterior samples: a non-significant result confirms unimodality &
posterior &
any &
\checkmark\\

Directional Bias of the Posterior &
A conjugate, symmetric prior and likelihood produce a symmetric posterior &
Difference of the mean and median of the posterior by at most 0.1*std.\ dev. Only done in Normal distributions &
posterior &
any &
\checkmark\\

Multi\-co\-linear\-ity &
Predictors are not pair-wise correlated &
Variance Inflation Factor (VIF) of each of the model's predictors: 5--10 means soft collinearity, $> 10$ is hard collinearity &
posterior &
any &
\checkmark\\

Data-Distribution match &
The observed values match the assumed likelihood distribution (e.g.\ continuous and unbounded for a normal distribution) &
Kolmogorov-Smirnov test between the distribution of the residuals and a Normal distribution &
likelihood &
linear &
\ding{55}\\

Information Content of the Likelihood &
The information of the likelihood produces a posterior that diverges from the prior &
KL divergence between posterior and prior: a value close to zero indicates the data carries little to no information &
likelihood &
any &
\checkmark\\

Categorical Variable Validation &
A variable encoded as categorical has discrete levels that are mutually exclusive and within the defined range of the data &
Direct checks for mutual exclusivity and exhaustiveness. Unique Value Ratio for discreteness: ratio close to 1 implies continuous variable &
likelihood &
cat. &
\ding{55}\\

Prior behavior on outcome scale &
After passing through the link function, a prior spreads its mass across the outcome scale rather than concentrate at its extremes. &
Evaluate the prior pushed through the link function and split into equal-width buckets over the likelihood. If every bucket holds at least a minimum share of the mass, the prior is sufficiently flat &
prior &
count, cat.&
\checkmark\\

K-1 model completeness &
A categorical model with K categories specifies exactly K-1 linear models, with one category held as the pivot &
Direct check of the parameters: compare size of vector of free parameters with expected cardinality derived from the data &
posterior &
cat.&
\ding{55}\\

Posterior predictive distributional validity &
Posterior predictive values stay within the support of the parameters in the data distribution &
Log-likelihood of the posterior draws: evaluate if there is an infinite value across draws &
posterior &
count, cat. &
\checkmark\\

Parameter reconstruction &
For data synthesized from known parameters, the posterior must contain said parameter values &
Evaluate if the original values are within the Highest Density Intervals (HDIs) of each parameter's posterior &
posterior &
any &
\checkmark\\

Link-function match &
The range of the link function is consistent with the domain of the parameters of the likelihood &
Sample the link function in question and check if they stay within the domain of the likelihood distribution &
prior &
count, cat. &
\checkmark\\

Over-Dispersion Variance Compliance &
The variance of the posterior does not exceed the theoretical variance of the likelihood distribution &
Direct comparison between the estimated variance of the posterior and the variance of the likelihood distribution &
posterior &
count&
\checkmark\\

Zero-inflation frequency &
The frequency of zeros in a Poisson outcome is in the credible interval of posterior predictive &
Evaluate if the observed amount of zeros is within the HDI of the posterior predictive distribution  &
posterior &
count&
\checkmark\\

Non-Centered Parameterization &
A grouping variable with low variation is non-centered, while one with high variation is centered &
Index of dispersion: evaluate if the variance-to-mean ratio of the non-centered variable is less or equal than 0.6 &
posterior &
hier.&
\checkmark\\

Shrinkage level variance &
In a hierarchical model, posterior estimates shrink toward the dataset mean (under full pooling),   toward each group's own mean (no pooling), and in-between the two (under partial pooling) &
The pooled and unpooled versions of the model are compiled and the spread of the group-level estimates is checked for the expected order &
posterior &
hier.&
\checkmark\\

Correlation matrix values &
In a multivariate model, the prior on the correlation matrix assigns low probability to high cor\-re\-la\-tions of varying effects &
The prior samples of the intercept--slope correlation are binned into three buckets across $-1$ and $1$. The prior passes if the central bucket holds the most mass. &
prior &
mult. &
\checkmark\\
Conditional Independence &
For independent variables, the changes of one do not affect the posterior of the other &
For each pair of independent variables, one is perturbed and both log-densities are re-evaluated across several random model states. The oracle is violated if the variable assumed independent responds &
posterior &
any &
\checkmark\\

\end{xltabular}

\vspace {-1ex}

\bibliographystyle {ACM-Reference-Format}
\bibliography {references}

\end {document}